\documentclass[10pt,journal,compsoc]{IEEEtran}

\usepackage{color}
\usepackage{graphicx}
\usepackage{hyperref}

\usepackage{amsthm}
\makeatletter
\def\th@plain{%
  \thm@notefont{}
  \itshape 
}
\def\th@definition{%
  \thm@notefont{}
  \normalfont 
}
\makeatother

\ifCLASSOPTIONcompsoc
  \usepackage[nocompress]{cite}
\else
  \usepackage{cite}
\fi

\ifCLASSINFOpdf
\else
\fi

\begin{document}

\IEEEoverridecommandlockouts
\IEEEpubid{\begin{minipage}{\textwidth}\ 
\centering\small{\copyright~2019 Personal use of this material is permitted.  Permission from IEEE must be obtained for all other uses, in any current or future media, including reprinting/republishing this material for advertising or promotional purposes, creating new collective works, for resale or redistribution to servers or lists, or reuse of any copyrighted component of this work in other works}
\end{minipage}}

\title{Privacy in Data Service Composition}

\author{Mahmoud Barhamgi, Charith Perera, Chia-Mu Yu, Djamal Benslimane,\\ David Camacho and Christine Bonnet
        
\IEEEcompsocitemizethanks{\IEEEcompsocthanksitem M. Barhamgi, D. Benslimane and C. Bonnet are with the Computer Science Department of Claude Bernard University, Lyon, France.\protect\\
E-mail: firstName.lastName@univ-lyon1.fr
\IEEEcompsocthanksitem C. Perera is with the Computer Science Department of Cardiff University.
E-mail: charith.perera@ieee.org
\IEEEcompsocthanksitem C. Yu is with the Computer Science Department of the National Chung Hsing University in Taiwan.
E-mail: chiamuyu@nchu.edu.tw
\IEEEcompsocthanksitem D. Camacho is with the Department of Computer Systems Engineering of Universidad Politecnica de Madrid.
E-mail: david.camacho@upm.es}
}

\IEEEtitleabstractindextext{%

\begin{abstract}
In modern information systems different information features, about the same individual, are often collected and managed by autonomous data collection services that may have different privacy policies. Answering many end-users' legitimate queries requires the integration of data from multiple such services. However, data integration is often hindered by the lack of a trusted entity, often called a \emph{mediator}, with which the services can share their data and delegate the enforcement of their privacy policies. In this paper, we propose a flexible privacy-preserving data integration approach for answering data integration queries without the need for a trusted mediator. In our approach, services are allowed to enforce their privacy policies locally. The mediator is considered to be untrusted, and only has  access to  encrypted information to allow it to link data subjects across the different services. Services, by virtue of a new privacy requirement, dubbed $k$-Protection, limiting privacy leaks, cannot infer information about the data held by each other. End-users, in turn, have access to privacy-sanitized data only. We evaluated our approach using an example and a real dataset  from the healthcare application domain. The results are promising from both the privacy preservation and the performance perspectives.
\end{abstract}

}

\maketitle

\IEEEdisplaynontitleabstractindextext

%
\IEEEpeerreviewmaketitle


\section{\textbf{Introduction}}

\IEEEPARstart{D}{ata} integration is the problem of bridging together a collection of data sources so that they can be queried as if they were parts of a single database. Despite  intensive research works devoted to that problem over the last few decades \cite{report,raj1}, developing a data integration system remains a challenging task. Data privacy, where the privacy of data subjects in one data source should be protected vis-a-vis  other sources and  the integration system as a whole, is among the key challenges involved in building a data integration system \cite{DustdarPST12,Barhamgcom}.

Most of existing multi-source data integration solutions are built as data warehouses where data is periodically collected from individual data sources and stored within a central data warehouse. Privacy is often approached by signing privacy agreements between data sources and the warehouse to specify who can access the data and for what purposes. However, such privacy agreements do not provide guarantees to individual data sources that their data would not be misused by the warehouse or any other stakeholder involved in the data integration system

In this paper, we explore an alternative multi-source data integration approach that gives data providers the control on their data while answering data integration queries and reduces the ability of the integration system to misuse data. We illustrate the research challenges addressed in this paper through a real-world example from the healthcare application domain.

\begin{figure*}[t]
	\centering
		\includegraphics[width=1.00\textwidth]{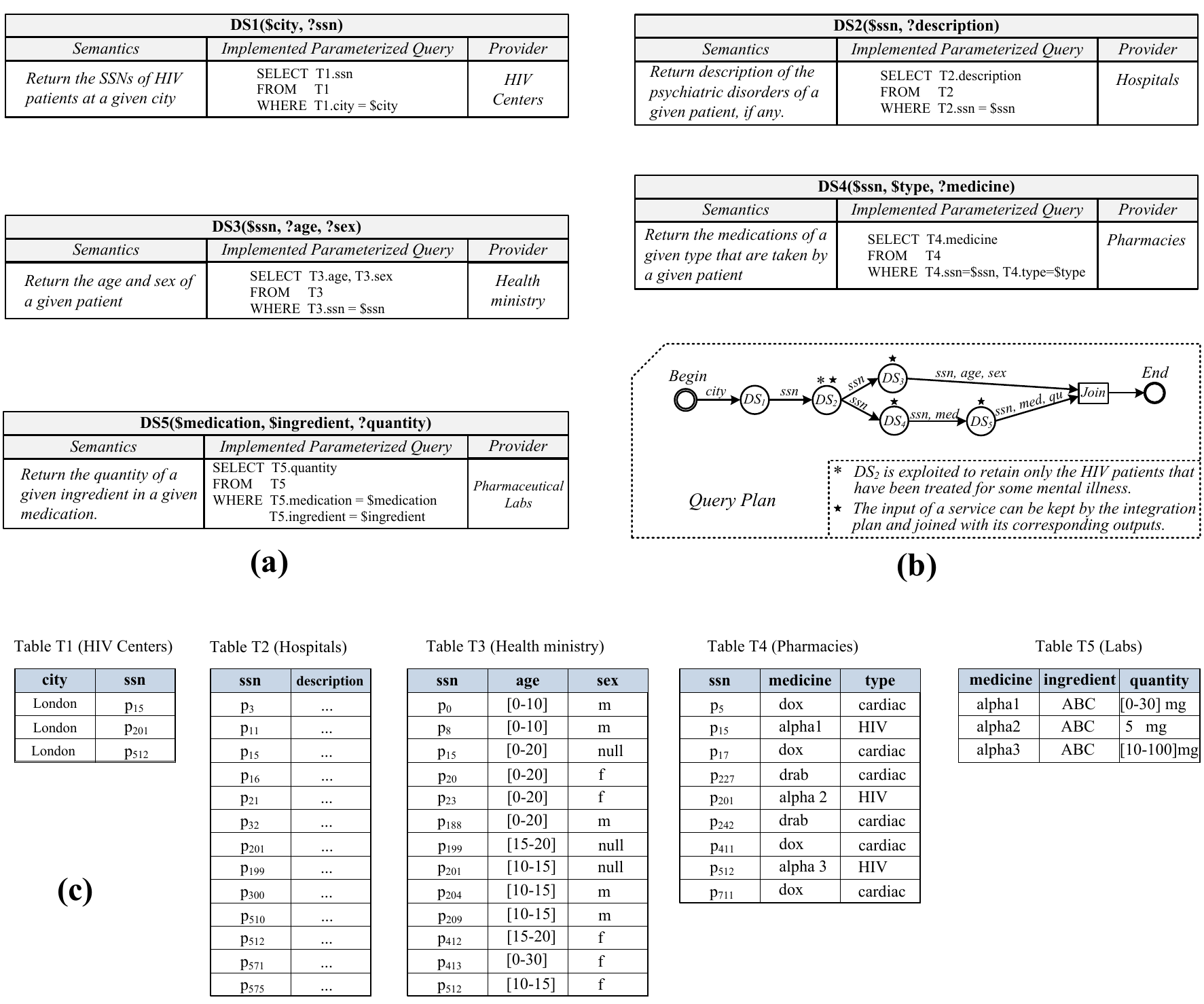}
	\caption{(a) The data  services of the running example; (b) A data integration plan; (c) Sample of the data accessed by the  services.}
	\label{fig:ScenarioComposition}
\end{figure*}

\vspace{-0.1cm}

\subsection{\textbf{Motivating Example}}

Data integration has important applications in the healthcare domain such as building the patient medical record and detecting the side effects of medications. Assume for example a data integration system with an access to the data services\footnote{It is a common practice in the healthcare domain to provide a service-oriented access to heterogeneous data sources \cite{Dogac12}. This class of services is known as \emph{data sharing services} \cite{DustdarPST12} or simply \emph{data services} \cite{CareyOP12}.} in Figure~\ref{fig:ScenarioComposition}-a, which in turn have access to the sample tables in Figure~\ref{fig:ScenarioComposition}-c. Assume we need to investigate the psychological side-effects of a specific ingredient of HIV medicines on female patients. Our sample data services can be composed as in Figure~\ref{fig:ScenarioComposition}-b to achieve our objective. Specifically, $DS_1$ is invoked with the desired {\tt city} to retrieve the identifiers (e.g. the social security numbers {\tt ssn}) of HIV patients. Then, $DS_2$, which is provided by psychiatric hospitals, is invoked with obtained {\tt ssn} numbers to retrieve the psychiatric disorders for which the patients have received some treatments for, if any. Then, for each HIV patient that has developed a psychiatric disorder, $DS_3$ and $DS_4$ are invoked in parallel to retrieve their {\tt age}, {\tt sex} and {\tt HIV medications}, respectively. Consequently, $DS_5$ is invoked to retrieve the quantity of the ingredient studied in each retrieved HIV medications. Then, the outputs of $DS_3$ and $DS_5$ are joined on {\tt ssn}.

\vspace{0.1cm}

The execution of the data integration plan (or also the service composition plan\footnote{We use the terms "service composition plan" and "data integration plan" interchangeably throughout the paper.}) in Figure~\ref{fig:ScenarioComposition}-b involves a challenging dilemma. That is, if individual services were to apply locally their privacy policies, and privacy-sanitize their output data then the plan cannot be executed. Note that in such case the social security number {\tt ssn}, which is used by the integration plan to link the different information features of the same patient, will not be disclosed by any of the services, as it is a personally identifiable information. On the other hand, if services share their output data with the integration system without any protection, then they may infer information about the data held by each other. For example, the provider of $DS_2$ may infer that his patients $p_{15}$, $p_{201}$ and $p_{512}$ (refer to table {\tt T2} in Figure~\ref{fig:ScenarioComposition}-c) are also AIDS patients if were provided with an access to the integration plan\footnote{This is a reasonable assumption since data providers may require to be informed how their data is exploited, by which entities and for what purposes. Often, this is defined in agreed upon privacy polices.}. He needs just to observe the inputs with which $DS_2$ is queried. Providers of $DS_3$ and $DS_4$ may also infer that those same patients have psychiatric disorders and AIDS. Moreover, the integration system (i.e. the entity responsible of coordinating the execution of the integration plan), as well as the end-user (i.e. the stakeholder studying the side-effects of AIDS medications) will learn the diseases, medications and ages of all patients transmitted by the services.

A naive solution to avoid information leakage among involved services would be by getting each of the services to ship a copy of its underlying data sources to a centralized data integration system, which can store received copies and answer queries locally. However, such solution has several drawbacks that make it impractical for real-world applications. For example, in application domains where the data is dynamic, i.e. updated frequently on the data provider side, the results computed using the copies in the centralized system may not include the latest updates on the providers' sides, which is crucial in critical domains such as the healthcare, where an incomplete or an erroneous query answer might have dramatic consequences on the patient's life. Moreover, such a centralized system may quickly become a target of choice for attackers due to the data concentration effect involved in retrieving copies from multiple data sources. Furthermore, current privacy regulations may prevent data providers from shipping the data they collect to a third-party which makes the solution undoable in practice.

In this paper, we propose a practical solution to allow the execution of multi-source data integration plans without leaking information about data subjects\footnote{The term data subject refers to the person whose data is collected and managed by data services (e.g. patients).} to involved stakeholders. That is, data services can not infer information about the data held by each other. Our solution can be used with trusted and semi-trusted (i.e. semi-honest) mediators. In the case of semi-trusted mediators, our solution enables involved data services to locally enforce their security and privacy policies (that relate to end-users), as the mediator cannot be trusted to enforce such policies. The solution still allows the mediator to join data subjects across the different services without leaking data from one service to another or to the mediator. In the case of trusted mediators, services can delegate the enforcement of their security and privacy policies to the mediator. The solution should just prevent data leakage among services.

\subsection{\textbf{\textbf{Existing Solutions}}}

Existing solutions for privacy-preserving data integration can be classified into three areas: \emph{privacy-preserving data publishing}, \emph{secure multi-party query computation} and \emph{trusted mediators based data integration}. We discuss below their limitations based on our running example.

\vspace{0.2cm}

\noindent\textbf{\textit{Privacy-preserving data publishing}}: Privacy models, such as $k$-Anonymity \cite{sweeney} and its variations $l$-Diversity \cite{Mach7} and $t$-Closeness \cite{LiLV07} compute an \emph{anonymized view} of a private table that can be shared with data consumers without the risk of disclosing the identity of specific data subjects (e.g. patients) in that view. Some recent works \cite{CFO13,D12} have attempted to apply the $k$-Anonymity concept (and its variations) in distributed settings where the anonymized view is computed using multiple private tables managed by autonomous data providers.

However, while $k$-Anonymity and its variations provide good privacy guarantees when the private table is owned by a single data provider, its implementations in distributed settings compromise the privacy of data subjects by leaking their private information to data providers while the latter compute the anonymized view. For example, all of the works \cite{CFO13,D12} assume that data providers, while computing the anonymized view, can know the list of data subjects they have in common\footnote{Data providers in these works exchange the identifiers of data subjects \textit{in clear} while they cooperatively compute the anonymized view.}, but none of them should know the specific attributes' values that are held by each other, beyond what is included in the computed view.  In our example, this means that the provider of $DS_2$ can learn that the patients $p_{15}$, $p_{201}$ and $p_{512}$ are also managed by the provider of $DS_1$ which is an HIV center, thus inferring that these patients have AIDS, thus  violating their privacy. Furthermore, such solutions are not convenient for data integration scenarios where data at the data providers' side is dynamic.

\vspace{0.2cm}

\noindent\textbf{\textit{Secure Multi-party Computation SMC}}: SMC protocols allow two or more parties to evaluate a function over their private data, while revealing only the answer and nothing else about each party's data. Initial solutions in this research area involved substantial computation costs that made them impractical \cite{mpc}. Several new solutions with an improved efficiency have emerged over the last few years \cite{Nayak15,WangAA14}. However, the efficiency improvements come at an expensive cost, most of the new solutions such \cite{Nayak15} require expensive (parallel) computation architectures that can be afforded by big corporations only. In addition, a recent evaluation study of the new solutions' efficiency \cite{kam} suggests that they can be applied to non-critical applications only, where end-users can accept delayed answers.

\vspace{0.2cm}

\noindent\textbf{\textit{Trusted mediators based data integration}}: Solutions in this category, such as \cite{YauY08,pairs}, rely on a centralized entity that can be trusted by all data providers to compute  data integration queries. For example, Yau et al. \cite{YauY08} present a privacy-preserving data repository that can collect, from each data provider, only the data necessary to compute the answer of a query. The data collected is hashed in the repository in such a way to prevent its reuse for computing other queries. However, while such solutions could be used when the integrated data is dynamic, they leak privacy sensitive information to data providers about the data held by each other.

\subsection{\textbf{\textbf{Proposed Approach and Contributions}}}

In this paper, we propose a practical multi-source data integration approach that would preserve the privacy of data subjects against the different stakeholders involved in answering a data integration query including data services, mediator and end-users. Our approach has the particularity of reducing data leakage among services to a practical amount that would prevent services from inferring sensitive information about data subjects. The mediator has access only to encrypted information to join data subjects across involved services. The proposed approach can be used for both trusted and semi-trusted mediators. 
Our solution can be exploited in numerous applications  where independent data sources should preserve the confidentiality of their data including healthcare \cite{Dogac12}, eGovernment \cite{report}, industrial collaboration scenarios \cite{raj2}, emergency management \cite{review1,LiuZYPPR17}, data management in smart environments \cite{review3,BarhamgiPGB18}, personal data markets \cite{IOTM}, multi sources data analytics \cite{rajiv1,review2}, etc.

\vspace{0.2cm}

\noindent The contributions of this paper are summarized as follows: 


\begin{itemize}

\item We define the main privacy requirements for the service-oriented class of  data integration. These requirements consider the different actors involved in a composition of services answering a query, including the \emph{constituent services} (i.e., data providers), the \emph{mediator} (i.e., the composition system) and the \emph{end-users} (i.e., data consumers).

\item We introduce a new privacy requirement, dubbed  \emph{$k$-Protection}, to ensure that there are no data leaks among services (i.e., data providers) during the query computation.  In a nutshell, when the output of a service $DS_i$ is used by a composition as input to another service $DS_j$, the  \emph{$k$-Protection} requirement protects the output of $DS_i$ by preventing $DS_j$ from distinguishing the exact input value from $k$ possible input values, where $k$ is set by $DS_i$. This would allow $DS_i$ to reduce the inference capability of $DS_j$ (and the other services in the composition) about its data below a practical threshold set by $DS_i$ itself.

	\item We propose an approach to evaluate multi-source queries over autonomous Web data services while respecting the $k$-protection requirement. 
	We validate our approach in the healthcare application domain by conducting a set of experiments using a real medical dataset of 1,150,000 records. The results show that our solution provides a practical privacy protection with acceptable performance overhead.
	
\end{itemize}

\vspace{0.1 cm}

The remainder of the paper is organized as follows. In Section 2, we define our privacy requirements with respect to services, the mediator, and the end-user. In Section 3, we present our approach for query evaluation. In Section 4, we evaluate the performance of our approach using a real medical dataset and discuss its applicability to real life application domains. In Section 5, we compare our approach with related research works and conclude in Section 6.


\section{Privacy Requirements in Service Composition}

In this section, we identify and discuss the data privacy requirements with respect to the different actors involved in a composition of services, including services, end-users and the mediator (or the composition system).

\subsection{Service Composition}

In this paper, we assume that data sources are exposed to the data sharing environment through Web APIs, i.e. Web services, to provide a standardized interface to data\footnote{The class of Web services that access and query data sources is known as \emph{Data Services} or \emph{Data Sharing Services} \cite{CareyOP12,DustdarPST12,YauY08}, and is motivated by the flexibility, and the interoperability that service oriented architectures could bring to data integration.}.

\begin{figure}[t]
	\centering
		\includegraphics[width=1.10\columnwidth]{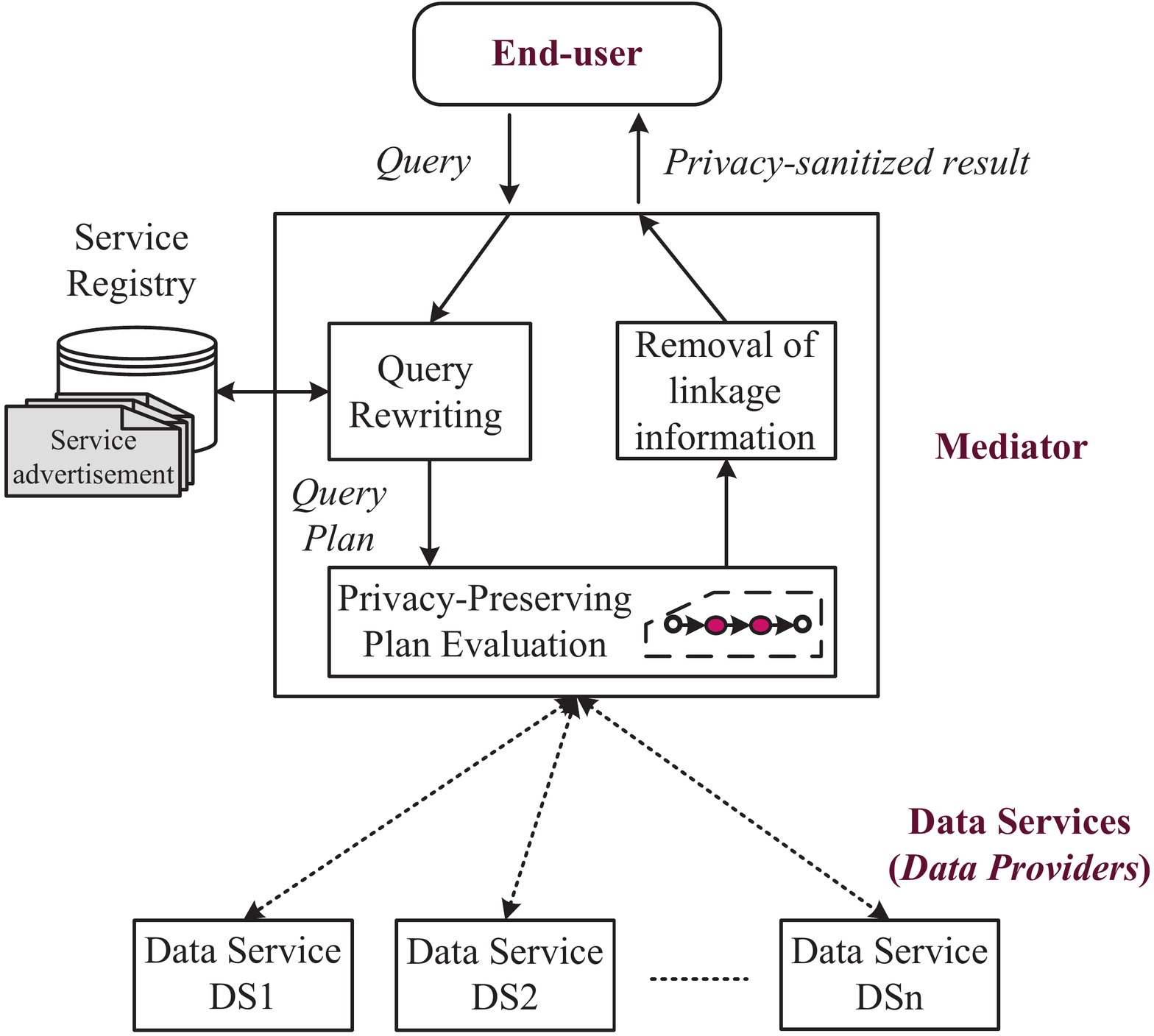}
	\caption{Privacy-Preserving Service Composition}
	\label{fig:System}
\end{figure}

End-users' queries are resolved by service composition as follows (Figure~\ref{fig:System}). Given a query and a set of available data services, the integration system compares the received query with the descriptions of available services to select the relevant ones. Then the integration system rewrites the query in terms of calls to selected services. The \emph{rewriting}, also called a composition, is then executed in such a way that preserves the data privacy relative to the different actors involved.

In this paper we will focus on the privacy issues raised by the execution of a composition. Readers interested in service selection and query reformulation in terms of services are referred to our previous work \cite{Barhamgi,BarhamgiEDBT} or to similar works \cite{widom06}.

In the following we formalize the notion of a composition  execution plan and define the requirements that should be satisfied in a privacy-preserving execution of a composition plan with respect to the different actors involved, including the mediator, end-users and involved data sharing services.

\vspace{0.2cm}

\noindent\textbf{Definition 1 (\textit{Composition Execution Plan})}: We adopt the definition given in \cite{widom06}. \textit{A composition execution plan is a directed acyclic graph $\mathcal{H}$ in which there is a node corresponding to each one of the data  services involved in answering the query, and there is a directed edge $e_{ij}$  from $DS_i$ to $DS_j$ if there is a precedence constraint between $DS_i$ and $DS_j$. We say that a service $DS_i$ must preceed  $DS_j$ if one of its outputs is an input for $DS_j$.}

The composition execution plan of our example is shown in Figure~\ref{fig:ScenarioComposition}-b. Please note that some nodes are simply preceded by end-users' input, e.g. $DS_1$. The query results can be simply computed by joining the outputs of services that are leaves in the plan (e.g. $DS_3$ and $DS_5$) \cite{widom06}.

\subsection{Data Privacy Requirements}

We focus on data privacy in this paper, i.e. the privacy of data subjects whose data is processed in a composition plan (e.g. patients in the running example), as opposed to the privacy of end-users who receive the final result which was adequately addressed in the literature.   

We say that the execution of  a composition  plan $\mathcal{H}$ is  privacy-preserving if it satisfies the following requirements with respect to the different actors involved including participating services, end-users and the mediator:

\subsubsection{Requirements with respect to services}

The execution of a composition should not leak information to its constituent services about the data held by each other. 

\vspace{0.1cm}

Let  $\Re(t_{x})$ represents the knowledge that a service $DS$ holds a given tuple $t_{x}$. Let $t_{x}$ concerns a data subject $x$ (e.g. a patient, a product, etc.). Let also $\Re_{ij}(t_{x})$ represents the knowledge leaked form $DS_i$ to $DS_j$  that $DS_i$ holds $t_{x}$. 

When a composition is executed without any privacy protection, the confidence of $DS_j$'s provider in $\Re_{ij}(t_{x})$ is $Pr_{j}(\Re_{ij}(t_{x}))$ = 1. For example, when $DS_3$ in the running example is invoked with the value $ssn$ = $p_{15}$, its provider will learn that both of $DS_1$ and $DS_2$ hold a tuple for the patient $p_{15}$, i.e. $Pr_{3}(\Re_{13}(t_{p_{15}}))$ = $Pr_{3}(\Re_{23}(t_{p_{15}}))$ = 1 (thus inferring   $p_{15}$ has HIV and mental disorders).

A service $DS_i$ can control the leakage of its data by keeping $Pr_{j}(\Re_{ij}(t_{x}))$ below an accepted threshold (fixed by $DS_i$). We define below a mechanism to control data leakage among services.

\vspace{0.2cm}

\noindent\textbf{Definition 2 (\textit{$k$-Protection})}:  \textit{Given a composition $\mathcal{H}$ and a vector $K= \{k_1, k_2, ..., k_n\}$, where $k_i$  is a positive integer determining the protection threshold (i.e. $1/k_i$) the service $DS_i$ must provide for its output tuples against the other services in $\mathcal{H}$, then for each edge $e_{ij}$ in $\mathcal{H}$,    $Pr_{j}(\Re_{ij}(t_{x}))$  must be $\leq$  $min(1/k_l)$, where $1/k_l$ is the protection threshold of  $DS_l$, which is, in turn, a (direct or indirect) parent of $DS_j$ in $\mathcal{H}$. Note that $DS_j$ has at least one parent in $\mathcal{H}$ (i.e. $DS_i$}, $k_l$ $\geq$ $k_i$).

\vspace{0.1cm}

The \textit{k}-protection mechanism is inspired by the concepts of \textit{k}-anonymity \cite{sweeney} and Private Information Retrieval PIR \cite{SionC07}.  Intuitively, this mechanism ensures that when a service $DS_j$ is invoked (with an input data value from $DS_i$), it must not be able to precisely determine  its input value between $k_l$  other possible input values (where  $k_l$ $\geq$ $k_i$). In other words, instead of invoking $DS_j$ with a precise input value $v$, $DS_j$ should be invoked with a generalized value of $v$ that matches with a range of values $V$ ($v$ $\in$ $V$) containing at least $k_l$ other possible values. This way, the certainty (i.e. the confidence) of $DS_j$ that $v$ is held by $DS_i$ is less than $1/k_l$. One possible way to implement this mechanism is to compute  $V$ such that it contains $k_l$ values for which $DS_j$ has an output. Please note that the \textit{k}-protection is similar to PIR in that when a service is queried (i.e. invoked) it does not know exactly the specific query it is executing on its own dataset.

\vspace{0.2cm}

\begin{figure}[t]
	\centering
		\includegraphics[width=1.00\columnwidth]{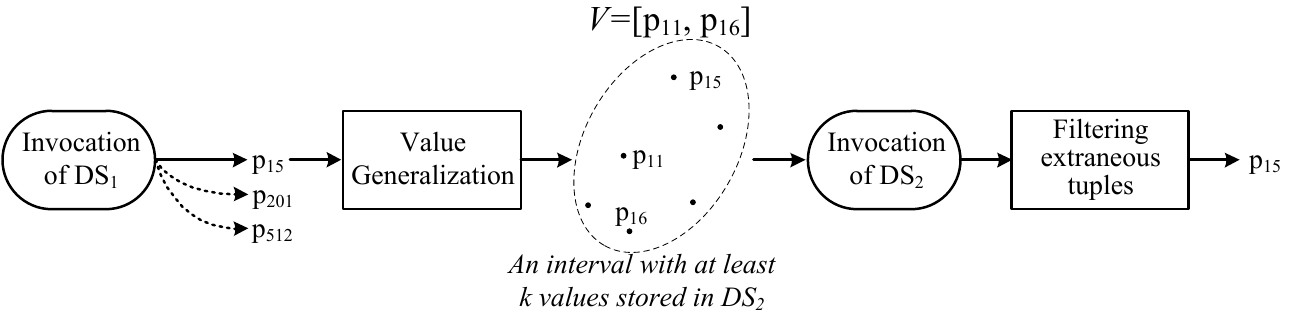}
			\caption{Applying the k-protection mechanism on the edge $e_{12}$ ($k$=3) }
	\label{fig:kprotection}
\end{figure}

\noindent\textbf{Example}: We continue on our running example. Given the data accessed by our sample services  in Figure~\ref{fig:ScenarioComposition}c, examples of the privacy breaches if these services were invoked without applying the \textit{k}-protection mechanism include: $DS_2$ will know that its patients {$p_{15}$, $p_{201}$ and $p_{512}$} have AIDS; $DS_3$ will know that these same patients, in addition to having AIDS, suffer  from severe psychiatric disorders, etc. Now, assume that  $k_1=3$, the \textit{k}-protection mechanism ensures that $DS_2$ must not be able to distinguish each of its input values (e.g. $p_{15}$) from at least 3 other values for which it has matching tuples in  its table \emph{T2}. Figure~\ref{fig:kprotection} shows how the \emph{k}-protection is enforced on the edge \emph{$e_{12}$}. The value $p_{15}$ is generalized into a range of values \emph{$V$} which contains at least three values (e.g. $p_{11}$, $p_{15}$ and $p_{16}$) for which $DS_2$ has matching tuples. After the invocation of  $DS_2$, the extraneous tuples are filtered out.

\vspace{0.2cm}

Preserving  data privacy against the different data providers involved in a query has not been addressed adequately in the literature. For example,  works on distributed privacy-preserving data publishing such as\cite{CFO13,D12} have focused on preserving the data privacy against final end-users while allowing involved data providers to know the list of data subjects they have in common. The $k$-Protection mechanism complements those works by making them immune to information leakage among data providers.

\subsubsection{Requirements with respect to end-users}

End-users are the entities that issue a query and receive the final result of executing the composition answering the query. Depending on the application considered, end-users could be trusted or untrusted by data subjects.  
For examples, care givers (e.g. primary care doctors, nurses, etc.) could be trusted to access (all or part of) the medical information of their patients for treatment purposes. Researchers who study the health conditions of a population, could be only trusted to access to anonymized data that cannot be linked to any specific patient.

End-users should be only allowed to access the information they are entitled to. This can be ensured either by applying privacy-aware access control policies of individual services in case of trusted end-users as in \cite{caise}, or by anonymizing the results returned by the composition to prevent the re-identification of data subjects.

\subsubsection{Requirements with respect to the mediator}

The mediator, i.e. the entity that executes the composition plan, is another important actor in a composition of services. It may not be necessarily managed by the final data recipient. It is responsible for carrying out the intermediary data operations in the composition plan (e.g. joining the outputs of different services, tuples selection, etc.).

Mediators could be trusted if managed by a trusted entity and are untrusted otherwise. An untrusted mediator should not have access to any Personally Identifiable Information PII. For example, the mediator should not be able to identify any of the patients $p_{15}$, $p_{201}$ and $p_{512}$ whose private data is circulating throughout the composition plan.

This implies that the data produced by a service should be protected by the service itself before being released to the mediator (and used for the invocation of other services). This can be ensured either by encrypting the released data (in case of trusted end-users) or by anonymizing it (in case of untrusted end-users).

\vspace{0.2cm}

Different applications may require different combinations of requirements, i.e. not all requirements should be respected at the same time. For example, in applications such as the healthcare, most often only the requirements related to services and end-users are relevant. In that domain, data integration is usually carried out by a trusted authority (e.g., a government agency) to, for example, discover new medical knowledge. In that case, it is important to protect data privacy against individual data providers (i.e. services) and end-users (e.g. researchers), whereas the mediator itself (i.e., the authority) is trusted. In application domains such as cybersecurity and terrorist fighting, often the requirement related to services is the most relevant. For example, consider a scenario where the objective is to proactively identify potential airplane terror attacks before they happen by identifying risky passengers (e.g., passengers with a criminal history and suspicious behaviors) on passenger lists of airline companies. Data providers in such case could be airline companies, police and intelligence services, banks, etc. The final end-user is a police inspector and the mediator is a governmental agency. In such case, the end-user and the mediator can be trusted, whereas data providers should not know they have a certain person in common, without a deep investigation from the inspector (e.g., an airline company should not know that one of its passengers has a criminal history before the case is fully investigated by the inspector).


\section{\textbf{A Privacy-Preserving  Query Evaluation Approach}}

In this section, we start by presenting our different assumptions and some key concepts. Then, we present 
an approach to evaluate multi-source queries while respecting the requirements discussed above.

\subsection{Context and Assumptions}
 In this work, we made the following assumptions. 
We consider a distributed environment with heterogeneous distribution of data. This means that different  services manage different features of information about the same set of data subjects. In contrast, in a homogeneous data distribution, different services manage the same features of information about different data subjects. The second case is easier to deal with since there is no real integration to be done. Therefore we only look at the first case.

 We consider a \textit{honest but curious environment}, where the stakeholders involved in the execution of a data integration plan will follow the given protocol. However, they may try to analyze exchanged data during the protocol execution. This setting is also known as \textit{semi-honest environment} in the literature \cite{EmekciAAG06}.

We assume that services can provide statistical information about their accessed datasets such as the service selectivity \cite{widom06}. The selectivity of a data service $DS$ relative to a range of input values $R$, denoted as $Se(DS, R)$, is the number of output values when $DS$ is queried with $R$. Let us consider the sample data in $T3$ (Figure~\ref{fig:ScenarioComposition}-c),  the selectivity of  $DS_3$ relative to some ranges is as follows: $Se(DS_3, [p_0, p_{1000}]) = 13$, where the range $[p_0, p_{1000}]$ includes the whole table $T3$; $Se(DS_3, [p_0, p_{10}]) = 2$, $Se(DS_3, [p_5, p_{20}]) = 3$.

\subsection{\textbf{An Overview of the Proposed Approach}}

    In our approach we assume that services, the mediator and end-users are independent entities. Our approach ensures the privacy requirements relative to those entities as follows.

    In our approach, data services involved in the integration plan can apply locally their security and privacy policies on non-identifier attributes. On the other hand, they are all required to encrypt the identifiers used by the mediator to join data by an Order Preserving Encryption Scheme OPES\cite{OPES}.  An OPES encrypts numeric data values while preserving the order relation between them. With OPES, we can apply equality and order comparison queries on encrypted data without decrypting the operands. Doing so, the mediator gets only access to encrypted values of identifier attributes and to values that are already protected by services for non-identifier attributes (e.g., by applying the desired anonymization techniques). This satisfies the privacy requirements relative to the mediator.  Once the integration plan has been executed, the mediator removes encrypted identifier attributes from result. This way, the final recipient will only have the anonymized data without any individually identifiable information. This satisfies the privacy requirements relative to end-users.

    To satisfy the privacy requirements relative to services, the mediator applies our $k$-Protection mechanism by generalizing the encrypted identifiers' values before using them to invoke the services. If a data service $DS_i$ was to be invoked with a value $x$ (originating from a parent service $DS_j$) and it is required  the certainty of $DS_i$ that $x$ is held by $DS_j$ to be  less than $1/k$, then $x$ is generalized through our protocol (presented in the following) to match with, at least, $k$ tuples in $DS_i$.  The value generalization is carried out by the mediator in collaboration with the service to be invoked.

Privacy requirements relative to end-users and the mediator are conventional requirements and have been extensively studied in the literature. For example,  the services can use any of the anonymization algorithms that implement the $k$-Anonymity and its variations \cite{sweeney,Mach7, LiLV07} to locally anonymize their data.  Privacy requirements with regard to services are, to the best of our knowledge, new, i.e., information leakage among data providers have not been property addressed in the literature. Therefore, in the following sections we focus on ensuring the $k$-Protection by presenting a practical protocol for generalizing encrypted identifier values.

The anonymization applied by the individual data services (and its utility) is out of the scope of this paper.  However, it is worth to mention that the anonymization can be realized by the services either \emph{in isolation} by directly applying one of the algorithms \cite{sweeney,Mach7, LiLV07},  or \emph{cooperatively} by extending one of the algorithms \cite{CFO13,D12} with our $k$-Protection mechanism presented in the next section.

\begin{figure}
	\centering
		\includegraphics[width=0.8\columnwidth]{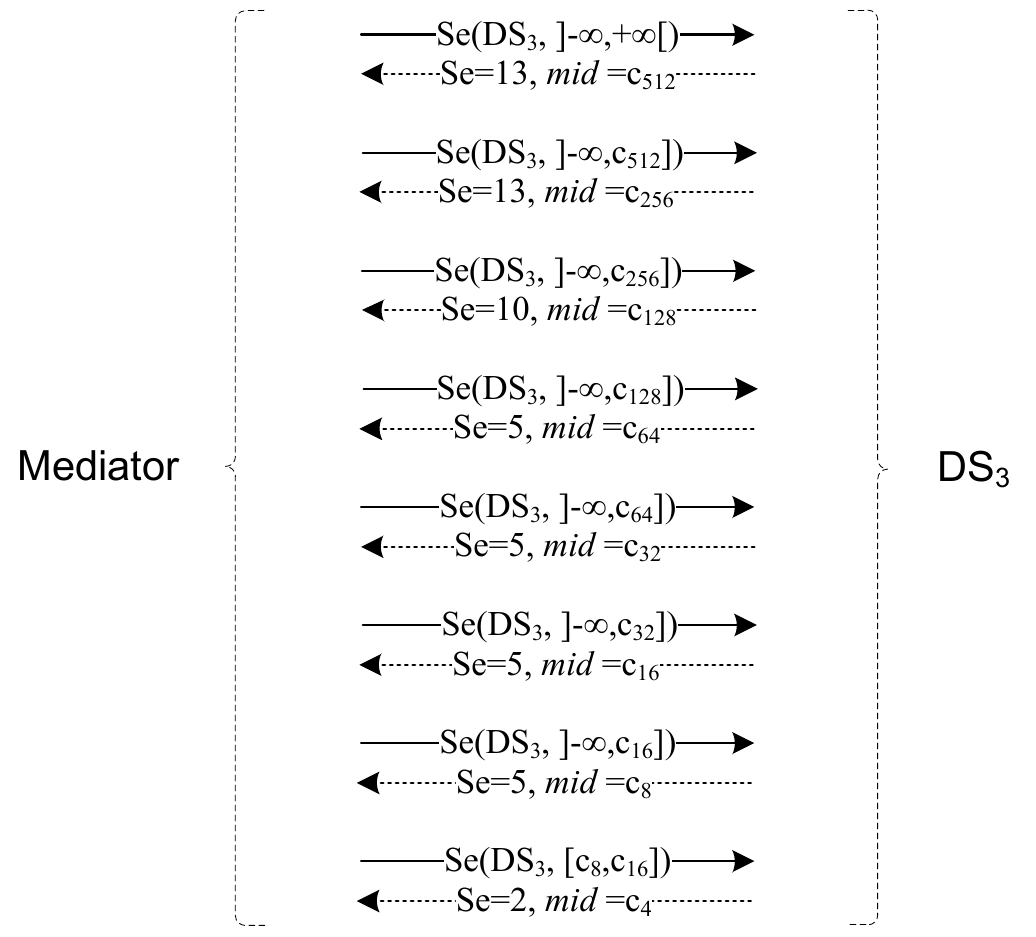}
	\caption{The communication rounds between the mediator and $DS_{3}$ to compute the generalized value.}
	\label{pro1}
\end{figure}

\subsection{\textbf{Generalization of  Encrypted Identifier Values}}

The objective of the generalization protocol can be formulated as follows: \emph{given an encrypted identifier value $x$ ($x$ $\in$ $X$) with which a service $DS_{i}$ should be invoked, allow the mediator to compute a generalized value\footnote{We assume that data services provide different operations (i.e., functions) allowing to query the underlying datasets by precise or generalized values (e.g., intervals).} $x'$ of $x$ such that $x'$ matches with, at least, $k$ possible values held by $DS_{i}$ where $k$ is maximum protection factor required by the parents of $DS_{i}$ in $\mathcal{H}$.} Note that the mediator cannot decrypt $x$ to generalize it alone. Rather, it needs to collaborate with the services to be invoked to carry out the generalization.

\vspace{0.1cm}

We first describe two naive approaches to generalize the identifier values and analyze their limitations. Then, we build on our analysis to define a hybrid approach that addresses the identified limitations.

\subsubsection{\textbf{Domain-based identifier generalization}}

A naive approach to compute $x'$ is to exploit the domain of the attribute $x$ (i.e., $X$). The idea here is to use $X$ as a starting value of $x'$ then to gradually increase its precision  by removing parts of $X$ until it is not possible to remove any part without violating  the $k$-Protection requirement. 

For this purpose, the mediator determines for each  data service $DS_i$  the protection factor \emph{k} that must be respected:   \emph{k} =  MAX($DS_j.k_j$), where $DS_j$ represents the parents of $DS_i$ in $\mathcal{H}$. Then, for each  input tuple \emph{t}, the mediator determines the minimum range of values \textit{R} [\textit{a}, \textit{b}] that should be used to invoke $DS_i$ instead of $t.x$. To this end, the mediator queries the selectivity of $DS_i$ with respect to a wide range of identifier values $R$ (we use the range ]-$\infty$, +$\infty$[ to denote the range covering all the tuples managed by $DS_i$) along with a value $mid$ occurring in the middle of the domain $X$. Then if the returned selectivity is greater than $k$, the mediator compares $t.x$ to $mid$ to determine the half of $R$ covering $t.x$. The last step is repeated with the obtained new interval until there is no interval with a selectivity greater than $k$. Then, $DS_i$ is invoked with the obtained interval. After the invocation, the mediator retains only the outputs related to $t.x$, i.e., the false-positives are removed by the mediator, as it knows the original encrypted input value $t.x$. 

\vspace{0.2cm}

\noindent\textbf{\textit{Example}}: Assume that the service $DS_3$ is to be invoked with the value $c_{15}$ (i.e., the encrypted value of $p_{15}$) and $k$ = 2. Figure~\ref{pro1} shows the messages exchanged between $DS_3$ and the mediator. For simplicity, the example assumes that $|X|$=1024.  First, the mediator queries the selectivity of $DS_3$ relative to a wide range of values (denoted by 
]-$\infty$, +$\infty$[). $DS_3$ replies that it has 13 values and the value $c_{512}$ is in the middle of  $X$. The mediator compares $c_{15}$ with $c_{512}$ and determine the new range ]-$\infty, c_{512}]$.  This step is repeated until the computed range cannot be divided while respecting the value of $k$.

 \vspace{0.2cm}

\noindent\textbf{\textit{Privacy Analysis}}: From a privacy perspective the protocol has the following limitations:

\begin{itemize}
	\item First, the mediator learns precise information about $X$, i.e., it receives the encrypted values of fixed data points in $X$ (i.e., $\frac{1}{2}X$, $\frac{1}{4}X$, $\frac{1}{8}X$...etc.). If the mediator knows the domain $X$, then it can map the encrypted values to their original values (since it knows their positions in $X$), or at least establish lower and upper bounds about each encrypted value.
	
	\item Second, it may leak additional knowledge to  services by violating the $k$-Protection requirement. For example, if the selectivity of the range $[c_{8}, c_{16}]$ in the previous example were 1 (instead of 2), then the mediator would return back to the range ]-$\infty, c_{16}]$. However, the service $DS_{3}$ would still know that the mediator is interested in the range $[c_{8}, c_{16}]$, as it has tested its selectivity. 
	One would think that  $|R|$ could be returned instead of the selectivity to avoid such a problem (by making sure that $|R|$ $\geq$ 2$k$ before splitting it), however the problem may persist as it could happen  that some values in the new range are not assigned to any individual, e.g., even though $|[c_{8}, c_{16}]|$ $>$ $k$, the values in that range, except for $c_{15}$, may not be assigned to patients (i.e., they are still virgin $ssn$ values), then $DS_{3}$ will infer that $x$ is $c_{15}$.

\end{itemize}

In addition, the protocol is not optimal. For instance, many of the rounds in Figure~\ref{pro1} (which involve important communication cost) did not reduce the selectivity, i.e., the first two ranges have the same selectivity, and the same applies to the fourth, the fifth, the sixth and the seventh range.

\begin{figure}[t]
	\centering
		\includegraphics[width=0.85\columnwidth]{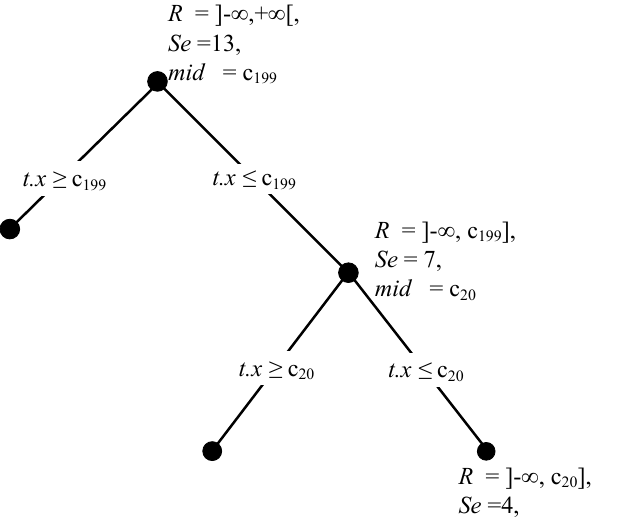}
	\caption{Generalizing the value $t.x$ = $c_{15}$ before invoking $DS_3$.}
	\label{fig:exempleKpro}
\end{figure}

\subsubsection{\textbf{Dataset-based identifier generalization}}

A second naive approach to generalize $x$  is to use the ordered dataset accessed by the service to be invoked, denoted by $S^{o}_{DS_{i}}$. In this approach, the mediator queries the selectivity of $DS_i$ with respect to a wide range of identifier values $R$ (we used the range ]-$\infty$, +$\infty$[ to denote the range covering $S^{o}_{DS_{i}}$) along with a value $mid$ occurring in the middle of $S^{o}_{DS_{i}}$. Then, if the returned selectivity is greater than $2k$, the mediator compares $t.x$ to $mid$ to determine the half of $R$ covering $t.x$. The last step is repeated with the obtained new interval until there is no interval with a selectivity greater than $k$. Then, $DS_i$ is invoked. After the invocation, the mediator retains only the outputs related to $t.x$, i.e., the false-positives are removed by the mediator, as it knows the original encrypted input value.

\vspace{0.1 cm}

\noindent\textit{\textbf{Example}}: We continue with our running example to show how we ensure the \emph{k}-protection requirement on the edge $e_{23}$. Assume that $DS_1$ and $DS_2$ require a protection factor \emph{k} = 3. The invocation of $DS_2$ returns the tuples corresponding to $c_{15}$, $c_{201}$ and $c_{512}$ where these values are the encrypted values of  $p_{15}$, $p_{201}$ and $p_{512}$. Instead of invoking $DS_3$ directly with the tuple $c_{15}$, the mediator generalizes $c_{15}$ as follows (refer to Figure \ref{fig:exempleKpro}). The mediator requests the selectivity of $DS_3$ with a range covering all its possibly managed values (i.e., $R$ = ]-$\infty$, +$\infty$[); $DS_3$ acknowledges it has 13 distinct values and that the value ($mid$ = $c_{199}$) occurs in the middle of these ordered values. The mediator compares $c_{15}$ to  $c_{199}$, and determines the new interval $R$= ]-$\infty$, $c_{199}$]. It then requests the selectivity of the new $R$ along with the new $mid$; the new values of $Se$ and $mid$ are 7 and $c_{20}$.   It determines again the new interval by comparing $c_{15}$ to  $c_{20}$. The new interval is $R$= ]-$\infty$, $c_{20}$] and its selectivity is 4. The algorithm stops here as if the new interval was divided then \emph{Se} will be less than \emph{k}.

\vspace{0.2 cm}

\begin{figure*}[t]
	\centering
		\includegraphics[width=1.00\textwidth]{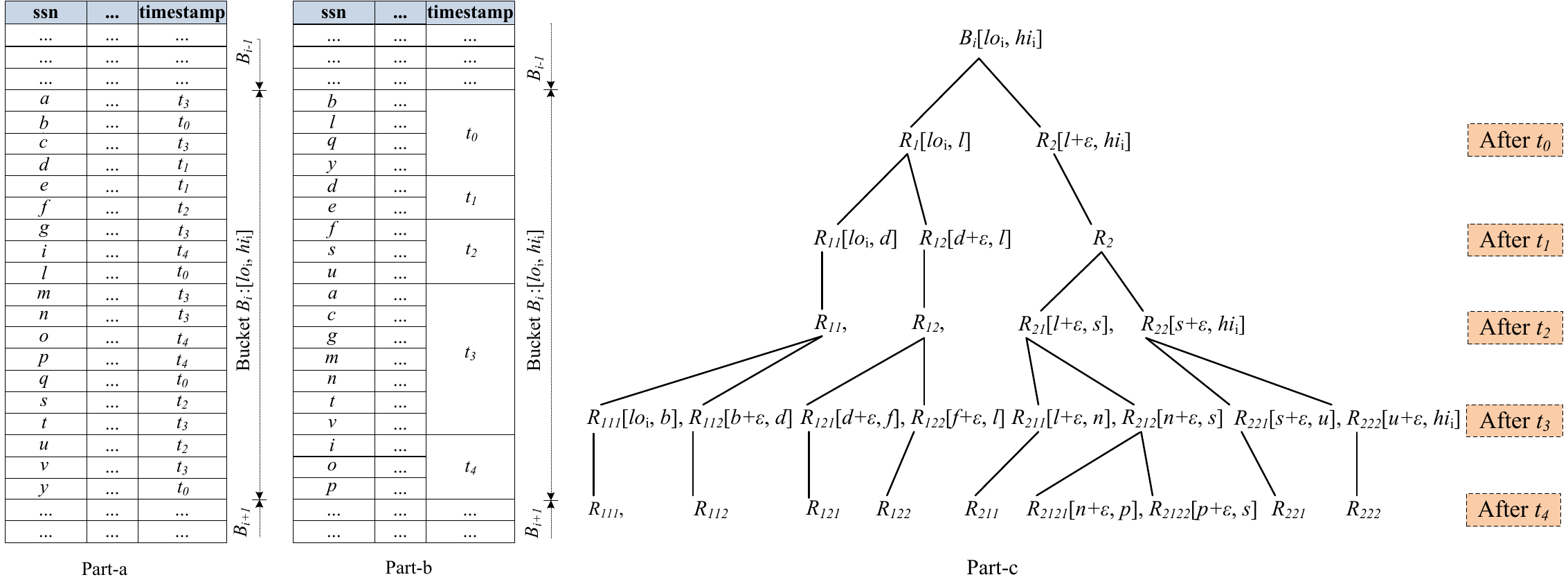}
	\caption{(a) a dataset partitioned into buckets; (b) data elements ordered by their time-stamps; (c) the computation of candidate ranges inside buckets.}
	\label{DBGenr}
\end{figure*}

\noindent\textbf{\textit{Privacy Analysis}}:  
This protocol does not suffer from the limitations discussed above. Specifically, it does not release the encrypted values of data points with known positions in $X$. It does not also test the selectivities of ranges that may contain less than $k$ values, as the protocol verifies that the selectivity of a range is $\geq$ 2$k$ before splitting it in two equal ranges.
xThe major limitation of this approach is that the boundaries of the computed range $x'$ depend on the dataset currently held by $DS_{i}$ (i.e., depend on $S^{o}_{DS_{i}}$), and may change if new tuples  were inserted in $S^{o}_{DS_{i}}$, or if some existing tuples were deleted, as we show in the following example. 

\vspace{0.1 cm}

\noindent \textbf{\textit{Example}}: Let us assume that  $S^{o}_{DS_{i}}$ = \{$a$, $b$, $c$, $f$\}, $k$ = 2, $x$ = $c$ and the computed $x'$ is [$c$, $f$]. If  the new tuples $d$ and $e$ were inserted, i.e.,  $S^{o}_{DS_{i}}$ = \{$a$, $b$, $c$, $d$, $e$, $f$\}, then $x'$ would become $x'$ = [$a$, $c$], and the provider of $DS_{i}$ would be able to infer that $x$ is $c$, as [$a$, $c$] $\cap$ [$c$, $f$] = \{$c$\}.

\subsubsection{\textbf{Hybrid protocol for identifier generalization}}

Based on the limitations discussed above, a good data generalization scheme should satisfy the following criteria. First, it should guarantee that the generalized value (i.e., the computed range $x'$) should always remain the same every time the composition plan is executed. In other words, the generalization scheme should be deterministic. Moreover, if the dataset $S^{o}_{DS_{i}}$ held by a service $DS_{i}$  is changed (because of data insertions or deletions), then the newly computed range should be a subset or superset of previously computed ranges for the same value $x$ and with, at least, $k$ intersecting values. Second, it should not leak additional information to the mediator that could help the latter to map the encrypted values to their real ones. 

\vspace{0.1 cm}

Before describing our proposed scheme for the generalization of encrypted data that avoids the discussed limitations, we first discuss two requirements that data services should satisfy to participate in the scheme. First, services must timestamp their accessed data. This requirement can be simply implemented by timestamping new data insertions. That is, when joining the data integration system, if the dataset accessed by a service is not time-stamped, the service can timestamp it with the current time, then timestamp new data insertions when they occur.  Second, the dataset accessed by a data service $DS_{i}$, denoted as $S^{o}_{DS_{i}}$, should keep track of the values of identifier attributes. That is, when a tuple inside the dataset must be deleted, the dataset should keep the value of the identifier attribute and its timestamp. 
These two requirements are realistic and can be easily implemented, as have been discussed in previous works in data integration such as \cite{ddd}.

We now present our data generalization scheme which satisfies the criteria discussed above by combining the two previously discussed data generalization approaches while avoiding their limitations. Our scheme proceeds along the following steps.

\begin{enumerate}

	\item The first step is carried out offline when  data  services join the data integration system. Every $DS_{i}$ partitions the domain $X$ to $m$ buckets. Services are free to choose the partitioning criteria and the number $m$. For example, a service may choose to divide $X$ into $m$ buckets with 50 stored values in each bucket, while another one may choose to divide $X$ into $m$ buckets with equal absolute length.
	
	\item The mediator executes the \emph{dataset based protocol} described above and narrows down the computed range $R$ as long as its  selectivity is $Se$ $\geq$ $\alpha$$\ast$$k$, where $\alpha$ is an integer value $>$ 1 selected by the mediator. As we will see later in our discussion, $\alpha$ guarantees that there is at least $\alpha$ candidate ranges satisfying the $k$-protection requirement in the $R$ computed so far.	We will explain the effect of $\alpha$ later  in our discussion. 
		
	\item  The service determines the bucket (or the set of buckets) that covers $R$ and divides them into ranges that respect the $k$-Protection requirement. Data elements in each bucket are organized into candidate ranges as follows: ($i$) they are ordered based on their time-stamps, i.e., data elements with the same time-stamp have the same order (which is the case for the initial set of data elements when $DS_{i}$ has joined the system). ($ii$) An initial set of candidate ranges are formed using the data elements with the highest order. ($iii$) Subsequent data points are inserted one after another in the computed ranges and when the selectivity of a candidate range becomes 2$k$ it gets split into two candidate ranges (that could also evolve independently).
			
	\item The computed ranges that intersect with the initial range $R$ (that is computed in the first step) are sent to the mediator, which can now select the candidate range covering $x$ and use it to invoke $DS_{i}$.			
	
\end{enumerate}

\begin{figure}[t]
	\centering
		\includegraphics[width=1.00\columnwidth]{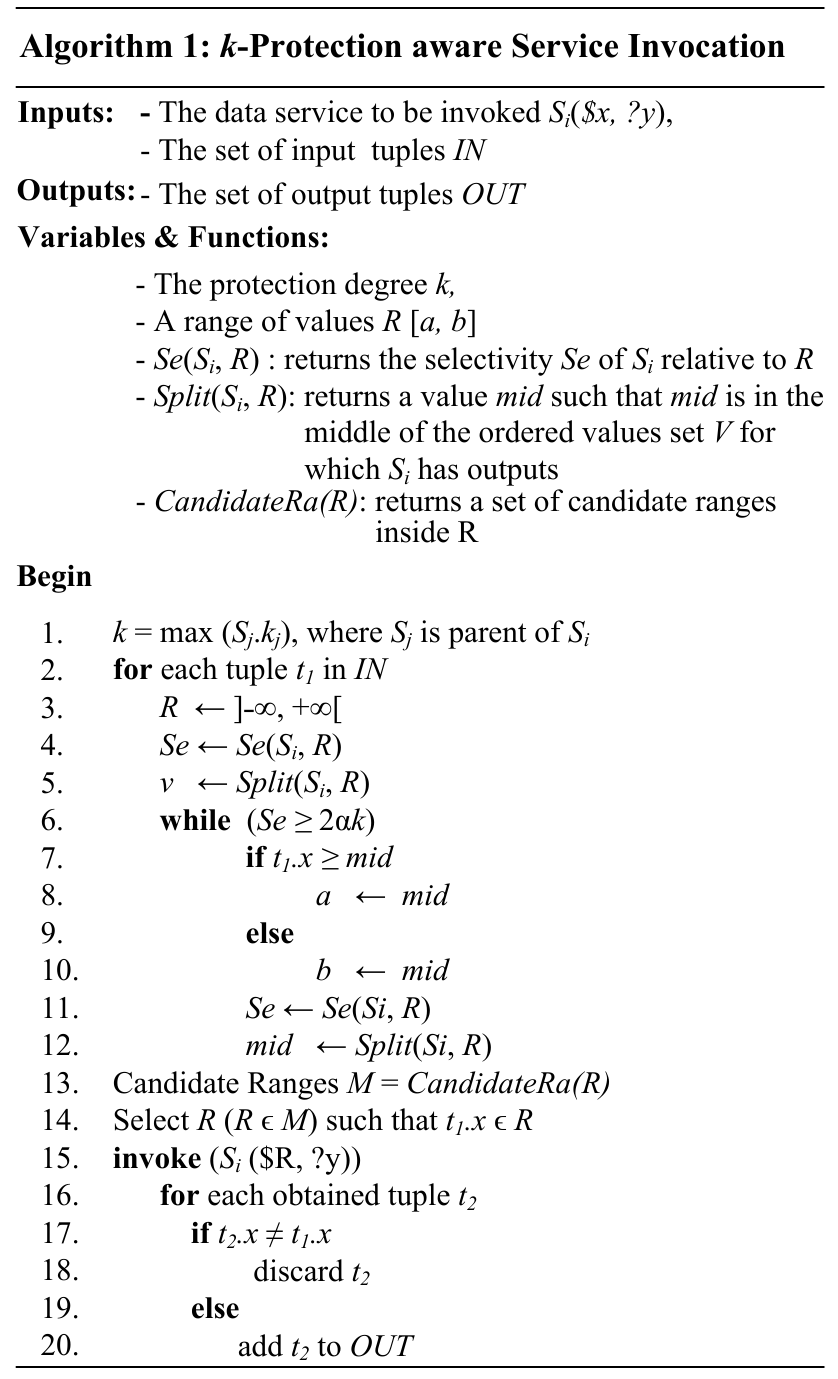}
	\label{fig:Algo}
\end{figure}

\noindent \textbf{\textit{Illustrative Example}}: Assume that the ordered dataset accessed by a data service $DS_{i}$ is shown in Figure~\ref{DBGenr} (part-a).  For simplicity, the figure shows the content of only one bucket (i.e., $B_{i}$). Each tuple in that dataset is time-stamped (e.g., the tuple $a$ has the time-stamp $t_{3}$). Assume also that $x$ = $q$, $k$ = 2, and $\alpha$ = 5. Assume also that after running the \emph{dataset based protocol} (in the second step) the computed range that satisfies the condition $Se(R) \geq  \alpha\ast k$ is $R$ =[$f$, $v$].

In the third step, the mediator asks $DS_{i}$ to compute the candidate ranges that cover $R$. To this end,  $DS_{i}$ finds the buckets that cover $R$ and computes their candidate ranges. As can be noticed from the dataset in Fig.~\ref{DBGenr} (part-a), $R$ $\subset$ $B_{i}$.

\begin{figure}[t]
	\centering
		\includegraphics[width=1.00\columnwidth]{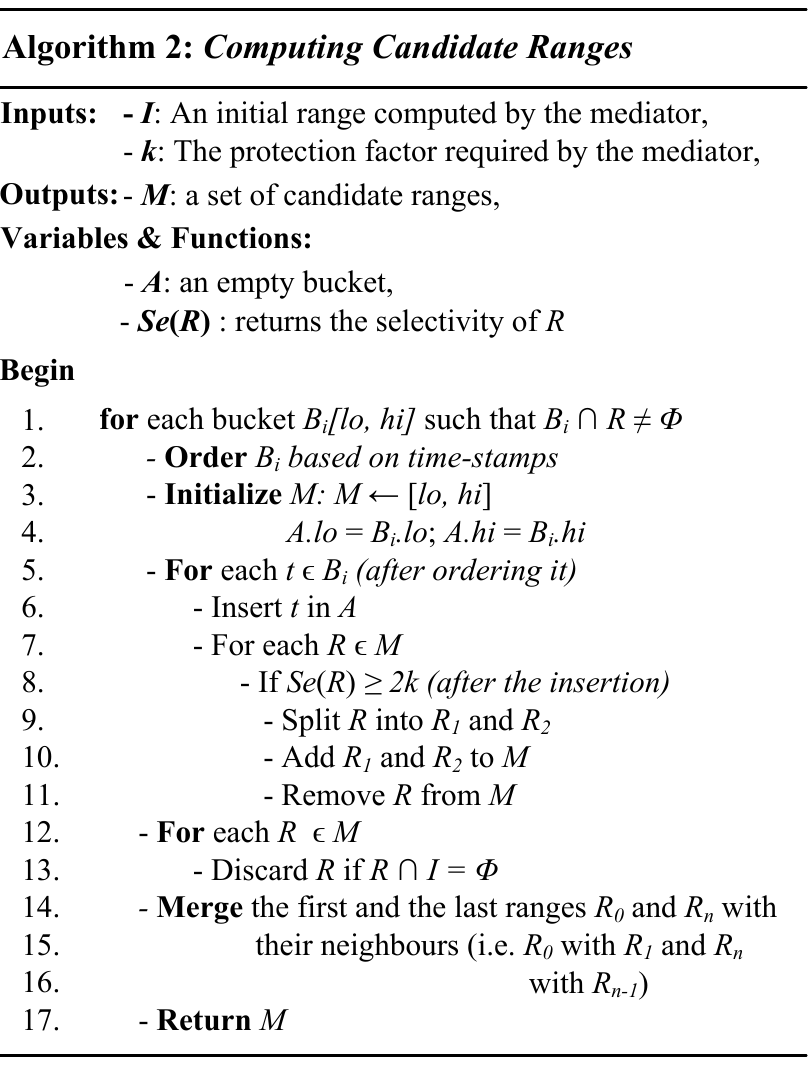}
	\label{fig:Algo}
\end{figure}

Figure~\ref{DBGenr} (part-b and part-c) shows how the ranges are computed. First, the service sorts the tuples inside $B_{i}$ based on their timestamps (part-b). Then, it starts to consider the tuples in $B_{i}$ one by one and divide $B_{i}$ into ranges with at least $k$ values. The tuples $b$, $l$, $q$ and $y$ are considered first (they have the same time-stamp $t_{0}$), which results into two initial ranges, i.e., $R_{1}$[$lo_{i}$, $l$] and $R_{2}$[$l$+$\epsilon$, $hi_{i}$], where $\epsilon$ $\approx$ 0. Then, the tuples $d$ and $e$ are considered (as they have the same time-stamp $t_{1}$). The selectivity of $R_{1}$ becomes 4 (i.e., 2$k$), and therefore it is split into two sub ranges $R_{11}$[$lo_{i}$, $d$] and $R_{12}$[$d$+$\epsilon$, $l$]. Similarly, when all the remaining tuples are considered, we have as a result 9 ranges (shown in Fig~\ref{DBGenr} (part-c).  

The range $R$ =[$f$, $v$] computed in the second step intersects with seven  ranges, i.e., $R_{121}$, $R_{122}$, $R_{211}$, $R_{2121}$, $R_{2122}$, $R_{221}$  and $R_{222}$. Since the range $R_{121}$  intersects partially with $R$, it is merged with $R_{122}$. The same applies to $R_{222}$, therefore it is merged with $R_{221}$. The obtained five candidate ranges are then sent back to the mediator, which will be able to select the final range with which the service will be invoked, i.e., $R_{2122}$ [$p$+$\epsilon$, $s$] ($q$ $\in$ $R_{2122}$).

Our proposed protocol is implemented by two algorithms, one is implemented by the mediator for computing the initial range R$\geq$ $\alpha$$\ast$$k$ and carrying out the service invocation (Algorithm 1) and one is implemented by the services for computing the candidate ranges (Algorithm 2), both  are self-described.

Note that the ranges computed inside each bucket are deterministic, i.e., the protocol results always in the same ranges whenever the composition plan is executed, as they are computed using the data insertion order in the dataset (which is deterministic). The addition of new data elements to the dataset does not violate the $k$-Protection requirement, as it only results in splitting a range into a set of ranges that intersect with it by at least $k$ elements. For example, when $x$ = $q$, if the plan was executed at an instant $t$ where  $t_{0}$ $\leq$ $t$ $\leq$ $t_{1}$ then the service will be invoked with range $R_2$, and when it is executed at another instant $t'$ where $t_{2}$ $\leq$ $t'$ $\leq$ $t_{3}$, then the computed range will be $R_{21} \subset R_{2}$. Since $R_{2}$ and $R_{21}$ intersect in $R_{21}$ which has at least $k$ values, i.e. the $k$-Protection requirement still holds.

\subsubsection{The effect of $\alpha$ on  privacy guarantee}

We intuitively explain the effect of $\alpha$ through an example. Consider the following dataset \{$a$, $b$, $c$, $d$, $e$, $f$, $g$\} and assume that $x$ = $d$ and $k$ is 2. 

When $\alpha$ = 1, then  the range $R$ that is computed in the second step (of our protocol) could take two different values   [$c$, $d$] and [$d$, $e$] (because of data changes) with equivalent probabilities $p$ = $\frac{1}{\alpha k}$ = $\frac{1}{2}$. When the same query is replayed, the probability of one of these two ranges happening if the other range has already happened is $p^2$ = $\frac{1}{(\alpha k)^2}$ = $\frac{1}{4}$.

When $\alpha$ = 2, then  the range $R$ that is computed in the second step could take four different values   [$a$, $d$], [$b$, $e$], [$c$, $f$] and [$d$, $g$]  with equivalent probabilities $p$ = $\frac{1}{\alpha k}$ = $\frac{1}{4}$. When the same query is replayed, the probability of  [$d$, $g$] happening when [$a$, $d$] has already happened (these two ranges intersect in $d$ ) is $p^2$ = $\frac{1}{(\alpha k)^2}$ = $\frac{1}{16}$.
Similarly, when $\alpha$ = 5, the probability of such privacy breach becomes $\frac{1}{(\alpha k)^2}$ = $\frac{1}{100}$, and when $\alpha$ = 10 this probability becomes $\frac{2.5}{1000}$.
In conclusion, practical values of $ \alpha$ (e.g., $\alpha$ $\geq$ 5) cut down the probability of privacy breaches to an accepted threshold. The value of   $\alpha$ can be computed as follows $\alpha$ = 
$\frac{\sqrt{\frac{1}{p}}}{k}$, where $p$ is the threshold of accepted probability.


\section{\textbf{Evaluation}}

\begin{figure*}[t]
	\centering
\includegraphics[width=1\textwidth]{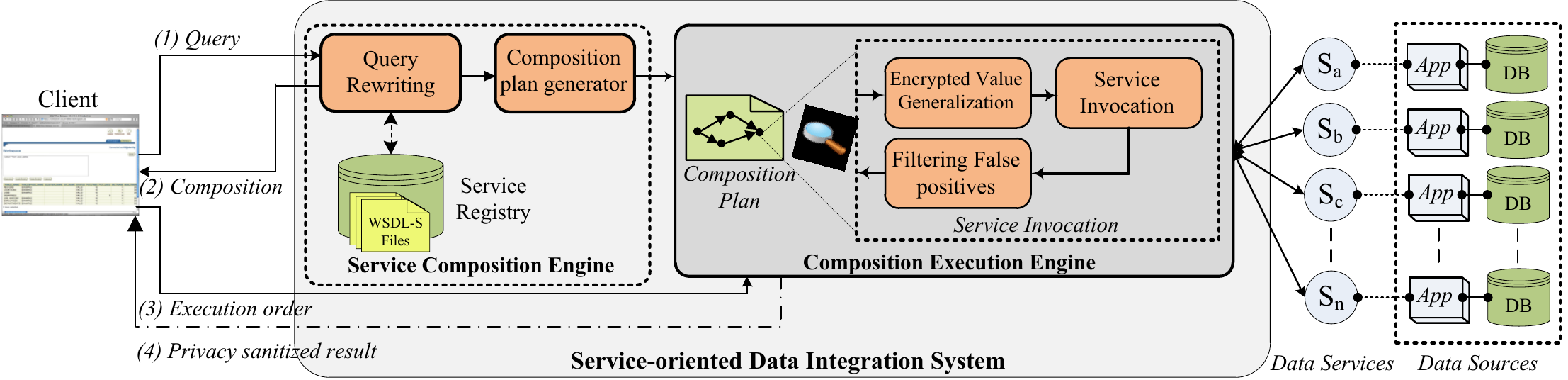}
\vspace{-0.3cm}
			\caption{The Evaluation Architecture}
	\label{fig:systemdemoWithoutRanking}
	\vspace{-0.3cm}
\end{figure*}

In this section, we present an evaluation study of our proposed approach, report on its performance and privacy preserving strength, and discuss some of the key applications where it can be used.

\subsection{\textbf{Evaluation Setup}}

To evaluate our approach, we used a real dataset that was provided to us by the European project  PAIRSE \cite{pairs}. The dataset was created by merging data from seven real databases of three French hospitals (specialized in psychiatry and cardiology).    
The dataset contained one big table with roughly 1,150,000 records (for approximately 850,000 patients). The table has the schema $R$($ssn$, $disease$, $dob$, $sex$, $city$, $physician$ $observations$). The $ssn$ values in the provided table were replaced by  synthetic numeric values by the original healthcare facilities. The same patient may have different rows in the table corresponding to the different diseases for which he or she has been treated.

We constructed three tables out of $R$: $R_1(ssn, disease, city)$, $R_2(ssn, disease)$ and  $R_3(ssn,$ $dob, $ $sex)$. $R_1$ contained only  heart patients (with a total number of 510,000 patients).  $R_2$ contained the patients that have been treated for a mental illness (with a total number of 403,000 patients).  For each of the tables $R_1$, $R_2$ and $R_3$, we constructed eight datasets of various sizes by randomly selecting patients of the tables. The constructed datasets have the sizes:  50K, 100K, 150K, 200K, 250K, 300K, 350K, 400K. For each dataset, we developed a data  service to access the dataset. These 24 services have the following three signatures: $DS_1(?ssn, ?disease, \$city)$, $DS_2(\$ssn, ?disease)$,  $DS_3(\$ssn,$ $?dob, ?sex)$, and were deployed on independent servers (with independent resources).

In our experiments we used the following composition:  $DS_1$$\circ$$DS_2$$\circ$$DS_3$; $DS_1$ is invoked with a given city name, then for each obtained patient $DS_2$ is invoked to verify if the patient has been treated for some mental illness, then $DS_3$ is invoked with only those who have been treated for a mental illness to retrieve their age and sex.

We implemented Algorithm 1 in Java and integrated it into the data integration system in \cite{pairs}. Figure \ref{fig:systemdemoWithoutRanking} shows the modified system, which consists of two main modules: \emph{service composition engine} and the \emph{composition execution engine}. The first module rewrites end-users queries into compositions of services. The second module executes the compositions. We modified, as the figure shows, the composition execution engine to accommodate Algorithm 1. All services are deployed on GlasFish 3.0 servers and datasets are stored in MySQL servers.  Algorithm 2 is implemented in Java on the server side and is accessed as a simple operation of each developed data service (i.e., each service provides an operation for querying its underlying dataset and a set of operations for participating in the proposed protocol, e.g., operations for computing the candidate ranges, computing the selectivity of a range, etc.).   

We conducted our experiments on machines with 3.2GHz Intel Processor running Windows 7 with 8 GB RAM.

\begin{figure*}[t]
	\centering
		\includegraphics[width=1.00\textwidth]{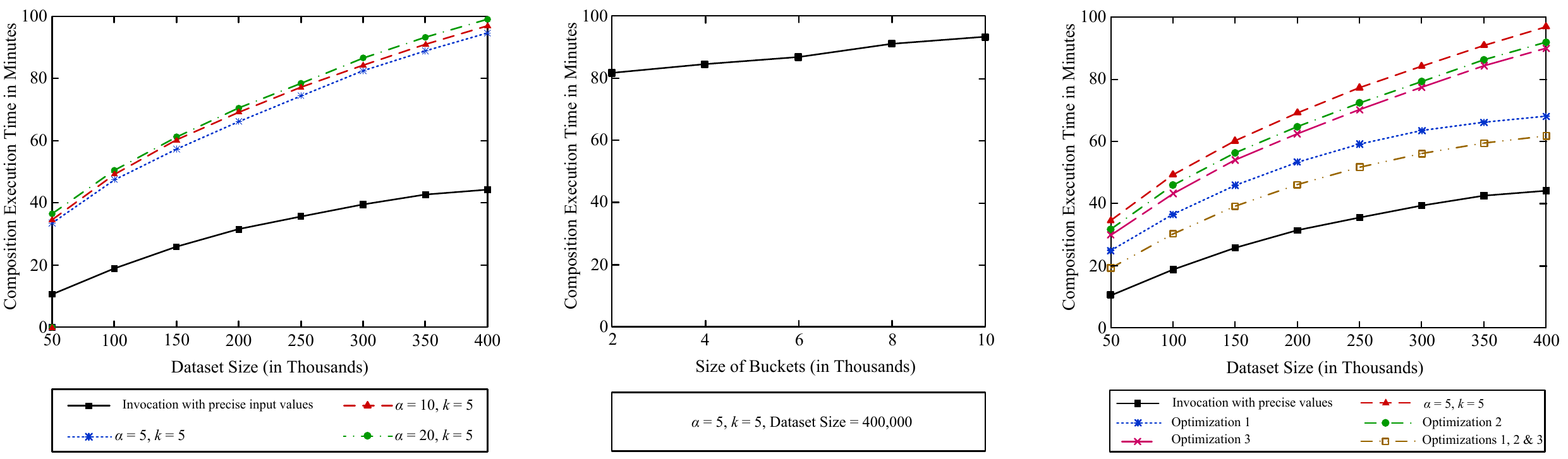}
	\caption{(a) The performance of the proposed approach as the size of datasets and the value of $\alpha$$\ast$$k$ increase; (b) The effect of the bucket size on performance; (c) The performance before and after applying the optimizations.}
	\label{fig:Experiments}
\end{figure*}

\subsection{\textbf{\textit{Performance Evaluation}}}

Assuming $n$ is the number of services in a composition, $l$ is the average number a service is invoked in a composition and $m$ is the average size of the datasets accessed by the services, the complexity of evaluating a query plan is of the order ~~~~~~~~~ $O$($n \ast l \ast $ $\log_2(\frac{m}{\alpha k}) \ast |Bucket|$).

We conducted a set of experiments to evaluate the performance of our privacy-preserving query evaluation approach using real  databases. Specifically, we evaluated the effects of ($i$) the protection factor $\beta$ = $\alpha$$\ast$$k$ and ($ii$) the size of accessed datasets on the composition execution time. 

We measured the composition execution time of our eight compositions;  all  compositions are identical, except that all services in each composition are accessing datasets of one of the aforementioned sizes.  We set the bucket size to 10,000 records (i.e., $|Bucket|$ = 10,000) for all services. We chose three high values for the protection factor $\beta$: $\beta_1$ = 25 ($\alpha$ = 5, $k$ = 5), $\beta_2$ = 50 ($\alpha$ = 10, $k$ = 5) and $\beta_3$ = 100 ($\alpha$ = 20, $k$ = 5). The probabilities of a privacy breach for these values are: $prob_1$ = $\frac{1}{(\beta_1)^2}$ = 0.0016, $prob_2$ = $\frac{1}{(\beta_2)^2}$ = 0.0004, $prob_3$ = $\frac{1}{(\beta_3)^2}$ = 0.0001.

The composition execution time was computed as follows. $R_1$ contained patients from 47 cities (with an average of 10,850 patients per city). Therefore, each composition was executed 100 times with each one of these 47 cities as input. Then, we computed the average execution time for each composition (i.e., for each one of the considered dataset sizes).   Figure \ref{fig:Experiments} (a) shows the obtained result together with the time required to execute the compositions without enforcing the $k$-protection requirement.

The obtained result shows that, for all the considered dataset sizes, the overhead involved in enforcing the $k$-protection requirement does not exceed two orders of magnitude of the time required to execute the composition without a protection. We view this as reasonable compared to the cost incurred by  private information retrieval and secure multi-party computation protocols  which exceeds by hundred times the cost of the original query without a protection. The reader is referred to \cite{kam} for discussion about the practicality of these approaches in real life applications. 

The result shows also that  increasing the values of  the protection factor $\beta$ = $\alpha$$\ast$$k$  only introduces  a minimal additional  overhead. This means that services can choose large values for $k$ (thus providing better privacy protection to their outputs) and the mediator can choose large values for $\alpha$ (thus a very small probability of a privacy breach) without degrading the overall performance.

Figure \ref{fig:Experiments} (b) shows the effect of changing the bucket size on the overall query evaluation time. Obviously, reducing the bucket size will reduce also the time required to compute the candidate ranges, and thus improve the overall performance. Please note that services are not required to use a large bucket size. It can be defined based on the accepted probability of a privacy breach. For example, if the bucket size was 2000, the probability of a privacy breach would be of the order   $\frac{1}{(2000)^2}$ = 0.00000025, which is a practical value. Additionally, the cost incurred in computing the candidate ranges inside a bucket can be offset altogether by computing these ranges offline, as we show in the next subsection.

\subsection{\textbf{\textit{Discussion}}}


As mentioned earlier, the value of $k$ is selected by each individual service to limit the inference capability of other services in the composition/query about the data held by the service. The parameter $\alpha$ is an integer value ($\alpha \geq$ 1) selected by the mediator to reduce the probability of a privacy breach (due to replay attacks) to a specific threshold. The higher the factor $\alpha*k$, the better the privacy protection. However, as the results show, the composition execution time increases also by increasing that factor.

The values of $k$ and $\alpha$ can be tuned to strike a balance between  privacy protection and  performance for the considered application domain. For example, the experiments show that if $\alpha$ = 10 and $k$ =5 ($\alpha * k$  = 50), the probability of inferring that a service holds a specific tuple is $\frac{1}{(\alpha * k)^2}$ = $\frac{1}{2500}$ = 0.0004 (i.e., the same query needs to be replayed at least 2500 times for that privacy breach to happen), and the composition execution time remains less than two times the time needed without any protection. In practice, the values of $k$ and $\alpha$ are selected such that the factor $(\alpha*k)^2$ is always higher than the number of times the same query can be executed by the same end-user (the system can limit the number a same query can be executed by the same end-user within a specific time window).

\subsection{\textbf{\textit{Performance Optimization}}}

The performance of our  approach can be improved further by considering the following  optimizations:

\vspace{0.2cm}

\noindent \textbf{\textit{Optimization 1: Reuse of pre-computed selectivities and ranges}}. At the composition execution time, the same service is likely to be invoked multiple times for different input values. The ranges and the selectivities that were computed in previous invocations (within the same composition execution occurrence) can be reused, even partially, instead of re-computing them each time the same service is invoked. Table 1 shows, for example, the numbers of patients returned by $DS_1$ when invoked with the cities ``Lyon 2",  ``Lyon 5" and ``Lyon 8", as well as the average number of the computed selectivities when  $DS_2$ is invoked. Without this optimization, the number of computed selectivities would be $log_2$(dataset size) = 18.6. The table shows also the numbers of reused candidate ranges.

\begin{table}[h]
\caption{Experimental Results}
\scriptsize
	\centering
		\begin{tabular}{|c|c|c|c|}\hline
		City  &  \# Patients (by $DS_1$) & \# Computed selectivities & \# Reused ranges\\\hline
		Lyon 2&13,738&7.3&257\\\hline
		Lyon 5&11,923&8.2&306\\\hline
		Lyon 8&9,736&8.7&119\\\hline
\end{tabular}
\label{cccc}
\end{table}

\vspace{0.1cm}

\noindent \textbf{\textit{Optimization 2: Use of user preferences}}. In real-life scenarios, many data subjects (e.g., patients, citizens, etc.) may accept to release some of their private data to some recipients for some legitimate data uses  (e.g., conducting medical research, law enforcement, etc.) without any protection. For example, the dataset $R$ was provided to us along with a patient preferences table specifying some of the entities each patient accepts to disclose his data; nearly 19\% of the patients accepted to share their medical data (including the $ssn$) with medical institutions for improving the healthcare system. In our second optimization, we exploit the preferences of data subjects to lift the privacy protection when a data subject consents to  data disclosure. For instance, in our running query the $k$-protection requirement could be lifted when  $DS_2$ and $DS_3$ are invoked with input values originating from those patients who accepted to disclose their data to medical institutions (such as those providing $DS_2$ and $DS_3$).

Please note that, one would think that invoking a service with a precise input value $x_{1}$ (of an individual who accepted to release her identifier) could invalidate the $k$-protection for another input value $x_2$ (of another individual who did not accept to release her identifier) if $x_{1}$ happens to be one of the $k$ possible values that match with the computed generalized value $x'_{2}$ (as the service would be able to eliminate one value of the $k$ possible values matching with $x'_{2}$). However, this will not happen in the first place, as the output tuples corresponding to $x_{1}$ will be retrieved when the service is invoked with $x'_{2}$, and thus there is no need to invoke it again. That is, lifting the k-protection requirement for those patients who agreed to release their identifiers has no impact on the  privacy protection provided for those who did not accept, as long as the services are invoked with input values of patients who did not agree to release their data before the ones who did.

\vspace{0.2cm}

\noindent \textbf{\textit{Optimization 3: Off-line computation of k-protection ranges}}. Services incur an important computation cost when they divide their buckets (that are relevant to an invocation) into ranges respecting the k-protection requirement (i.e., Step 3 of the proposed hybrid protocol).  However, this step can be carried out offline in an incremental fashion, as they insert new tuples.

We conducted a set of experiments to evaluate the performance improvement that could  result from the  optimizations  presented above.  Figure \ref{fig:Experiments} (c), shows the obtained results when ($i$) none, ($ii$) only one and ($iii$) all of the optimizations is/are applied. The result shows that applying those optimizations substantially improves the performance. In fact, the query execution becomes less than two orders of magnitude of the query execution time when no protection is applied (regardless of the considered dataset size).

\subsection{\textbf{\textit{Privacy Analysis}}}

Our protocol is immune against replay attacks. That is, even if the same composition was executed several times with the same inputs, data service providers cannot increase their confidence sufficiently to infer with certainty the identity of a data subject with which they are queried. For example, as  explained when we analyzed the effect of $\alpha$ on the privacy guarantee, if $\alpha$ was set to 20, and $k$ was set to 5, then the probability of a privacy breach happening is $prob$ = $\frac{1}{(\alpha k)^2}$ = $\frac{1}{10000}$; and when $\alpha$ = 200 and $k$ = 5, that probability becomes 1000000. That is, the same composition needs to be executed million times in order to have two candidate ranges (R=$\alpha$$\ast$$k$) whose intersection is the identifier of a targeted data subject.

Some would argue that the order preserving encryption scheme (OPES) of Agrawal et al. \cite{OPES} is insecure and can be broken by the mediator. However, for us, that scheme is only a means, not an end. We can use any of the recent OPESs reviewed in \cite{CO12} or even a partial implementation of the Homomorphic encryption \cite{GentrySW13}. Our protocol requires only to allow the mediator to carry out equality and comparison operations (i.e., =, $<$, $>$) which could be realized with any of the aforementioned schemes.

One limitation of the approach is that when the domain $X$ of the identifiers is small, if the mediator knows all of the values in $X$, then it can, after a certain number of queries (i.e., compositions), build a mapping table between the encrypted values and the real ones. However, this can be overcome by encrypting the identifiers with a different key every time a query is executed. 
Moreover, the proposed generalization protocol itself can be applied on both real and encrypted values. In some application domains, the mediator would be trusted, and service providers may not need to encrypt the identifiers when they are released to the mediator, whereas the mediator would still need to generalize the identifiers (with our protocol) when it queries individual services.


\section{Related Work}

Our solution for privacy preservation relates closely to research works in the areas of \emph{mediator based data integration}, \emph{privacy-preserving data publishing}  and \emph{secure multi-party computation}. In this section, we review some of the most prominent works in those areas and compare them to our approach.

Several research works addressed the privacy issue in the context of data service composition. For example, Yau et al. in \cite{YauY08} proposed a repository to answer end-user queries by integrating data across autonomous data sharing services. The privacy in that work is addressed by ($i$) allowing the repository to collect only the data necessary to compute the result of a query (instead of retrieving the whole datasets behind the services), and by ($ii$) hashing the identifiers (used to link data subjects) between the services of each couple of interconnected services in the composition graph. Unfortunately, this solution does not resolve the privacy breaches addressed in our work. In fact, the repository still has access to intermediate and integrated data, as the hashing is carried out by the repository itself. In addition, services can learn information about the data held by each other, as they are invoked with precise data values. Benslimane et al. in \cite{pairs} proposed a privacy-preserving access control model to preserve the data privacy against data consumers. Queries are rewritten, by a mediator, to include applicable privacy constraints before they are resolved by service composition. Ammar and Bertino in \cite{Amar14} proposed to take into account the context of data consumers when the privacy policies are applied on their queries before they are resolved by service composition. Unfortunately, these solutions, while providing good privacy protection against data consumers, do not provide any privacy protection against un-trusted mediators and data services.

Our data generalization protocol for service invocation with imprecise values (e.g. ranges) is reminiscent of the works on the Private Information Retrieval (PIR) problem  \cite{WangAA14,SionC07}. The objective of these works is to execute private queries on a remote server without letting the server to learn anything about  executed queries or their results. While current PIR protocols could provide strong privacy guarantees against untrusted services, they involve prohibitive computation costs (mostly due to their cryptographic nature) that make them impractical \cite{SionC07} for real applications. In contrast, our model for service invocation takes a practical stance on the performance/information leakage trade-off; i.e.  services are allowed to learn a controlled  amount of knowledge in return for an important performance improvement.

Several research works have addressed the problem of distributed data integration \cite{D12,X09}. Fung and Mohammed \cite{D12} proposed a data mashup system that can anonymize data from several data sources to provide data consumers with datasets that satisfy the \emph{k}-anonymity property. While that work was geared towards data consumers, our focus is on data providers and the integration system itself (i.e. the mashup server). We assumed in our work that data providers can freely apply their privacy policies (e.g. data anonymization techniques) on their sides. However, our solution can be extended with that of \cite{D12} to make the anonymization on the integration system side. Jurczyk and Xiong \cite{X09} proposed an algorithm to securely integrate horizontally partitioned data from multiple data sources.  However, that work does not address the vertically partitioned data which is closer to our work.

\section{Conclusion and future work}

In this paper, we proposed a privacy-preserving approach to evaluate data integration queries over autonomous data services. Our approach allows services involved in a query to apply  their privacy policies locally. The data integration system (i.e. the mediator) is given  encrypted information only to allow it to link data subjects across the different services which, in turn, cannot learn information about the data held by each other. We evaluated our approach in the healthcare application domain, and the results showed that our solution can be applied to cost effectively integrate voluminous datasets. We intend to extend our approach to improve the performance further. An interesting direction to explore is the possibility of invoking services with chunks  of input data tuples \cite{widom06}. The data value generalization algorithm should then be extended then to generalize chunks instead of single data items.

\ifCLASSOPTIONcaptionsoff
  \newpage
\fi



\bibliographystyle{IEEEtran}
\bibliography{IEEEabrv,ref}
%




%






\end{document}